\documentclass[12pt,english]{article}

\usepackage{geometry}
\usepackage[titletoc,title]{appendix}

\usepackage{setspace}
\usepackage{sectsty}
\allsectionsfont{\normalsize\raggedright\centering}

\usepackage[]{tocloft}
\addtocontents{toc}{\cftpagenumbersoff{section}}
\addtocontents{toc}{\cftpagenumbersoff{subsection}}

\makeatletter
\renewcommand{\@cftmaketoctitle}{}
\makeatother

\usepackage{amsmath, amssymb, mathptmx, comment, tensor, enumitem, epigraph, etoolbox, relsize, setspace}

\usepackage[textsize=tiny]{todonotes}

\usepackage{natbib}
\usepackage{url}
\makeatletter
\def\url@leostyle{%
    \def\UrlFont{\sf}}{\def\UrlFont{\small\ttfamily}}
\makeatother
\bibpunct
{(} %the opening bracket symbol
{)} %the closing bracket symbol
{,} %the punctuation between multiple citations
{a} %the letter `n' for numerical style, or `s' for numerical superscript style, any other letter for author-year
{} %the punctuation that comes between the author names and the year
{;} %the punctuation that comes between years or numbers when common author lists are suppressed

\usepackage{authblk}

\usepackage{graphicx}
\usepackage{babel}

\usepackage{amsmath}
\usepackage{amsfonts}

\usepackage{framed}

\usepackage{xifthen}


\numberwithin{equation}{section}

\usepackage{txfonts}
\usepackage[T1]{fontenc}
\newcommand{\citetbjps}[2][]{\ifthenelse{\equal{#1}{}}{\citeauthor{#2} ([\citeyear{#2}])}{\citeauthor{#2} ([\citeyear{#2}], #1)}}
\newcommand{\citealtbjps}[2][]{\ifthenelse{\equal{#1}{}}{\citeauthor{#2} [\citeyear{#2}]}{\citeauthor{#2} [\citeyear{#2}], #1}}
\newcommand{\citepbjps}[2][]{\ifthenelse{\equal{#1}{}}{(\citeauthor{#2} [\citeyear{#2}])}{(\citeauthor{#2} [\citeyear{#2}], #1)}}
\newcommand{\citeyearbjps}[2][]{\ifthenelse{\equal{#1}{}}{[\citeyear{#2}]}{[\citeyear{#2}], #1}}
\newcommand{\citeyearparbjps}[2][]{\ifthenelse{\equal{#1}{}}{([\citeyear{#2}])}{([\citeyear{#2}], #1)}}
\newcommand{\citeposbjps}[2][]{\ifthenelse{\equal{#1}{}}{\citeauthor{#2}'s ([\citeyear{#2}])}{\citeauthor{#2}'s ([\citeyear{#2}], #1)}}

\begin{document}

\title{The local validity of special relativity from a scale-relative perspective}
\author{Niels Linnemann, James Read and Nicholas J.~Teh}
\date{}

\maketitle

%% no headers on the first page:
\thispagestyle{empty}

\begin{abstract}   
    Most contemporary physicists hold that the local validity of special relativity (SR) within general relativity (GR) is expressed by means of an interdependent cluster of mathematical concepts, one of which is the existence of normal coordinate systems. Nonetheless, there remains conceptual work to be done with regard to this `standard story' on the local validity of SR in (a) clarifying how a network of mathematical concepts is recruited in a particular modelling context in order to account for the local validity of SR within GR, and (b) highlighting the richness and subtlety of this mode of modelling, as well as the way in which it interacts with the concept of `approximate Killing symmetry'. With this paper, we carry out this work, thereby also defending the standard story from concerns recently voiced in the philosophy of physics literature.
\end{abstract}

\tableofcontents

\section{Introduction}

\label{sec:intro}

Conventional wisdom has it that general relativistic physics is effectively special relativistic for `sufficiently small neighbourhoods.'\footnote{We include scare quotes here because, of course, `sufficiently small neighbourhoods' is a term of art which, ultimately, will require unpacking.} This position is reflected in leading textbook presentations of the matter (often in relation to principle(s) of equivalence)---consider, for example, this passage from Steven Weinberg's seminal textbook on gravitation:
\begin{quote}
The Principle of Equivalence rests on the equality of gravitational and inertial mass, demonstrated by Galileo, Huygens, Newton, Bessel, and Eötvös [...] Einstein reflected that, as a consequence, no external static homogeneous gravitational field could be detected in a freely falling elevator, for the observers, their test bodies, and the elevator itself would respond to the field with the same acceleration [... But e]ven the observers in Einstein's freely falling elevator would in principle be able to detect the earth's field, because objects in the elevator would be falling radially toward the center of the earth [...] Although inertial forces do not exactly cancel gravitational forces for freely falling systems in an inhomogeneous or time-dependent gravitational field, we can still expect an approximate cancellation if we restrict our attention to such a small region of space and time that the field changes very little over the region. Therefore we formulate the equivalence principle as the statement that \textit{at every space-time point in an arbitrary gravitational field it is possible to choose a ``locally inertial coordinate system" such that, within a sufficiently small region of the point in question, the laws of nature take the same form as in unaccelerated Cartesian coordinate systems in the absense of gravitation}. There is a little vagueness here about what we mean by ``the same form as in unaccelerated Cartesian coordinate systems," so to avoid any possible ambiguity we can specify that by this we mean the form given to the laws of nature by special relativity [...] There is also a question of how small is ``sufficiently small." Roughly speaking, we mean that the region must be small enough so that the gravitational field is sensibly constant throughout it, but we cannot be more precise until we learn how to represent the gravitational field mathematically. \citepbjps[p.~68, emphasis in original]{WeinbergFirstEdition}
\end{quote} 

So, according to Weinberg, within `a sufficiently small region', laws in general relativity (GR) take the same form as in special relativity (SR), i.e., as in the inertial frames of SR. (It follows that Weinberg is also saying that the physical effects modelled by the mathematical concept of `spacetime curvature' can be neglected in the right regime. We will return to that later on.)
In fact, this is arguably exactly what Einstein himself had in mind---or so he writes in a letter to Painlevé:\footnote{The following translation from \citetbjps[p.~135]{lehmkuhl2021equivalence}.}
\begin{quote}
According to the special theory of relativity the coordinates
$x, y, z, t$ are directly measurable via clocks at rest with respect to the coordinate system. Thus, the invariant $ds$, which is defined via the equation $ds^2 = dt^2-dx^2-dy^2-dz^2$, likewise corresponds to a measurement result. The general theory of relativity rests entirely on the premise that each infinitesimal line element of the spacetime manifold physically behaves like the four-dimensional manifold of the special theory of relativity. Thus, there are infinitesimal coordinate systems (inertial systems) with the help of which the $ds$ are to be defined exactly like in the special theory of relativity. The general theory of relativity stands or falls with this interpretation of $ds$. It depends on the latter just as much as Gauss’ infinitesimal geometry of surfaces depends on the premise that an infinitesimal surface element behaves metrically like a flat surface element [...] . \citepbjps{EinsteinPainleve}
\end{quote}

Let us call what is expressed in the Weinberg and Einstein quotes above the `standard story' regarding the effective emergence of SR within GR. Although Weinberg's above presentation of the standard story leaves room for explication---as he readily admits---we think that it makes eminent sense. All that remains is to carry out these explications, and articulate clearly the way in which the standard story is baked into a sensible interpretation of the geometric formalism of GR.\footnote{By `geometric formalism' we are not taking a stance in the dynamical/geometric debate (on which see e.g.~\citepbjps{BrownRead}). In the following, `geometry' is to be thought of in terms of mathematics---it is \emph{prima facie} not any less `dynamical' to talk in terms of a geometrical \emph{formalism}.}

Our view of the standard story should be uncontroversial for the physics community. Indeed, most contemporary physicists hold that the local validity of SR within GR is expressed by means of an interdependent cluster of mathematical concepts, one of which is the existence of normal coordinate systems, to which the above quotes allude.
Nonetheless, there remains conceptual work to be done in (a) clarifying how a network of mathematical concepts is recruited in a particular modelling context in order to account for the local validity of SR within GR, and (b) highlighting the richness and subtlety of this mode of modelling, as well as the way in which it interacts with the concept of `approximate Killing symmetry'.\footnote{To be clear: in this article we are concerned primarily with making sense of the local validity of SR, \emph{given} the completed theory of GR. Of course, one can---as Einstein did---use the local validity of SR as an heuristic towards the development of GR (see \citepbjps{Einstein1916}). This heuristic aspect, although interesting, is not our primary focus in this piece---though for discussion (which we take to be complimentary to our own), see \citepbjps{HetzroniRead}.}

The structure of the article is this. In \S\ref{sec:loc1}, we explain how the standard story regarding the local validity of SR in GR is cashed out in terms of the geometric formalism of GR.
In this section, the formal characterization of SR is geometric (in terms of Minkowski spacetime); however, as is well-known, we can also characterize SR in terms of the Poincar\'{e} symmetries of Minkowski spacetime---thus, one can also pose the effective emergence of SR from GR in terms of the effective emergence of Poincar\'{e} symmetry---or, in terms of approximate Poincar\'{e} symmetry.\footnote{There is a precedent for discussing these issues in the recent philosophy literature: see \citepbjps{read2018two}, \citepbjps{FletcherApprox}.}
We thus take up the matter of a \textit{symmetry}-based understanding of the `effective' emergence of SR in \S\ref{sec:localvalidity2}. Since the description of SR with which we are concerned is an `effective' one (i.e.,~one that is approximately valid in a certain regime, relative to the background theory of GR), one must here confront a difficult question: how should one articulate the relevant notion of `approximation' in `approximate symmetry'? In this section, we argue that \citetbjps{Harte} has provided a comprehensive and clear solution.
%---at least in the limited case that we will consider (i.e. the dynamics that is baked into standard ways of interpreting the geometric formalism of GR).
We then show that this symmetry-based understanding dovetails nicely with the account of the standard story which we gave in \S\ref{sec:loc1}.
Finally, we consider in \S\ref{sec:response} how our work interacts with other recent investigations into the local validity of SR in GR---in particular, those of \citetbjps{fletcher2022local} and \citetbjps{gomes}.

\section{The local validity of special relativity from normal coordinates}

\label{sec:loc1}

In this section, we first present the standard story regarding the local validity of SR in GR in terms of normal coordinates (\S\ref{sec:normalcoordinates}). We then explain how this sense in which GR is locally special relativistic is best understood from a scale-relative perspective (\S\ref{sec:EFTlikeaspect}).

\subsection{Normal coordinates}

\label{sec:normalcoordinates}

In the context of classical spacetime theories formulated in terms of Lorentzian manifolds, normal coordinates are coordinates systems which are defined on a normal neighbourhood\footnote{The normal neighbourhood $\mathcal{N}(p)$ relative to a point $p$ is the set of points that are uniquely connected to $p$ by geodesics \citepbjps[p.~42]{poisson2011motion}.
%COMMENT_1
The normal neighbourhood relative to \emph{set} of points $S$ is the union of normal neigbourhoods $\mathcal{N}(p)$ for each point $p \in S$, i.e. $\mathcal{N}(S) = \cup_{p \in S} \mathcal{N}(p)$.} relative to an embedded submanifold $S$ with vanishing intrinsic curvature such that, on $S$, they are formally indistinguishable from a Lorentzian frame in Minkowski spacetime (i.e.,~a frame in which the metric field diagonalises).\footnote{Strictly speaking, these are `orthonormal' (as opposed to just `normal') coordinates---i.e.,~the coordinates in which the metric diagonalises (orthonormality) and in which the connection coefficients of the associated compatible derivative operator vanish (normality). In the course of this article we will generally mean `orthonormal normal' by `normal' coordinates. See \citepbjps{Knox2013-KNOESG} for philosophical discussion of this distinction, and \citepbjps{Iliev} for more technical background.} For example, Riemann normal coordinates (RNCs) are defined on a normal neighbourhood relative to a sub-neighbourhood which is a \emph{point}; Fermi normal coordinates (FNCs) are defined on a normal neighbourhood relative to a sub-neighbourhood which is a \emph{line}, and so forth. (We will be more precise about the definitions of these different forms of normal coordinates in what follows.)

In this subsection, we emphasize that RNCs are most helpfully thought of in terms of \textit{invariant} geodesic deviation structure. To this end, we first introduce the notions of (i) a (manifold-)coordinate-independent tetrad lab frame (\S\ref{sec:labframe}), and that of (ii) Synge's world function, which codifies geodesic deviation structure (\S\ref{sec:worldfunction}). Having done so, we will be in a position to show that (iii) RNCs are nothing but projections of the (coordinate-independent) world function onto the (coordinate-independent) tetrad `legs' (i.e., the four vector fields which comprise the tetrad) (\S\ref{sec:riemannnormalcoordinates}). We end by commenting on some other normal coordinate systems (\S\ref{sec:ferminormalcoordinates}).

\subsubsection{A coordinate-independent lab frame}

\label{sec:labframe}

In this article, we use Greek indices $\mu, \nu, \ldots$ for coordinate indices on the manifold, capital Latin letters $I, J, \ldots$ for coordinate indices on the tangent space, and lowercase Latin indices $a,b,\ldots$ for abstract indices on the manifold. Consider then the point $x'$---called a `base point' in what follows---with normal neighbourhood $\mathcal{N}(x')$ and a certain set of coordinates $\{z^{\mu}\}$ defined on that normal neighbourhood.\footnote{Our presentation here largely follows that of \citetbjps{poisson2011motion}.} (We follow the convention whereby the base point with respect to which a given normal coordinate system is defined is denoted with a prime.) Relative to the base point $x'$, define an orthonormal `lab' frame $e\indices{_{I}^{\mu}}$ such that
\begin{equation}\label{eq1}
g_{\mu \nu} e\indices{_{I}^{\mu}} e\indices{_{J}^{\nu}} (x')= \eta_{IJ} (x'), % \qquad \frac{D e\indices{_{I}^{\mu}}}{d \lambda} = 0,
\end{equation}
where $\eta_{IJ} = \text{diag}(-1, 1, 1, 1)$ is the Minkowski metric. Here and in what follows, we follow the convention whereby the coordinate values of geometric objects at the base point are denoted by primed indices---so, \eqref{eq1} can equivalently be written as
\begin{equation}
g_{\mu' \nu'} e\indices{_{I}^{\mu'}} e\indices{_{J}^{\nu'}} = \eta_{IJ}. % \qquad \frac{D e\indices{_{I}^{\mu}}}{d \lambda} = 0,
\end{equation}
The so-called dual tetrads are defined as $e\indices{^I_{\mu'}} := \eta^{IJ} g_{\mu' \nu'} e\indices{^{\nu'}_J}$; thereby, $g_{\mu' \nu'} = \eta_{IJ} e\indices{^I_{\mu'}} e\indices{^J_{\nu'}}$. 

As the lab frame is not specific to a special set of coordinates $\{z^{\mu}\}$, we can talk of a (manifold-)coordinate-independent tetrad $e\indices{_{I}^{a'}}$ where $a'$ is now an abstract index on the manifold. The existence of such a frame is special in GR insofar as it is adapted to the canonical tangent space of GR.\footnote{It has been questioned whether the right tangent space for specific GR spacetimes is always the four-dimensional tangent space isomorphic to $\mathbb{R}^n$ \citepbjps{mukhanov}. However, these considerations in fact go beyond GR; one generalised theory allowing for other tangent spaces is Einstein-Cartan theory (see, for instance, \citepbjps{EinsteinCartan}).}

At this stage, as we are apparently dealing with a lab frame defined at a `point', it is worth contemplating what is meant by a `point' in GR to begin with. Notably, a manifold point $p \in M$ does not represent an extensionless event---rather, it can be understood as a mathematical notion which we bring to bear upon some particular modelling context. The context which will concern us in this article is that in which the characteristic length scale of some object (a black hole, a lump of matter, etc.) is sufficiently small relative to some other relevant background length scale that the object is well-modelled by a (mathematical) point. Thus understood, the point is rather analogous to point particles.

\subsubsection{The world function}

\label{sec:worldfunction}

For any $x \in \mathcal{N}(x')$ there is a unique geodesic linking $x$ to $x'$; denote the geodesic by $z^{\mu}(\lambda)$ with $\lambda$ the affine parameter ranging from $\lambda_0$ to $\lambda_1$ such that $z(\lambda_0) = x'$ and $z(\lambda_1) = x$. Now, the world function relative to the base point $x'$ and its neighbouring point $x$ is defined as 
\begin{equation}
\label{eq:world}
\sigma(x, x') = \frac{1}{2} (\lambda_1 - \lambda_0) \int_{\lambda_0}^{\lambda_1} g_{\mu \nu}(z(\lambda)) t^{\mu} t^{\nu} d\lambda,
\end{equation}
with $t^{\mu} := \frac{d z^{\mu}}{d \lambda}$ tangent to the geodesic.  Note that the world function evaluates to $ \epsilon (\lambda_1 - \lambda_0)^2 /2$, where $\epsilon := g_{\mu \nu} t^{\mu} t^{\nu} = \text{const}$ across the geodesic, with $\epsilon = \pm 1$ in the spacelike/timelike case, and $0$ for the null case.\footnote{Further properties of the world function are described by \citetbjps{poisson2011motion}.}
Again, just like the tetrad frame, the world function is a coordinate-independent expression relative to the manifold.

The world function is originally due to Synge (and thus also known as `Synge's world function'; see \citepbjps{Synge} for his exposition). Standard use cases include (i) geodesic triangles in spaces of small curvature, (ii) the definition of spatial lengths, and (iii) studying the effect of curvature in the measure of angles (see \citepbjps{Synge}, \citepbjps{Cardin}, and references therein). Of central relevance for its significance in any case is the relationship between its covariant derivative and the exponential map: $\sigma^{a'}(x, x'):=\nabla^{a'} \sigma(x, x')$ gives (minus) the tangent vector of the unique geodesic between $x'$ and $x$ emanating from $x'$; the associated function $\sigma^{a'}_{x'}: \mathcal{N}(x') \subset M \rightarrow T_{x'} M, x \mapsto \sigma^{a'}(x, x')$ is then the inverse to the exponential map (defined here as $U \subset T_{x'} M \rightarrow \mathcal{N}(x') \subset M, v \mapsto \gamma_v (1)$ where $v$ is a tangent vector at $x'$ with parameterization $\lambda$, and $\gamma_v (1)$ the unique point after parameter distance $1$). 

Given its intimate relation to the exponential map, it is not surprising that the world function can be said to encode geodesic structure in various ways: the world function $\sigma(x, x')$ is tantamount to a Lagrangian; it thus encodes---in combination with Hamilton's principle (which requires that the action determined by the Lagrangian be stationary)---the geodesic equations \emph{qua} equations of motions \citepbjps{Ballmann}.
Similarly, the world function can be seen to encode information regarding geodesic flow: upon Legendre transformation of the action, the corresponding equations of motion are the generators of cogeodesic flow of the tangent bundle \citepbjps{Cardin}.\footnote{See \citepbjps{Uppsala} for a pedagogical account of the Riemannian case.}

In addition, the derivative of the world function at base point $x'$, seen as a function $\sigma^{a'}_{x'}(x)$ of $x$, encodes geodesic deviation structure, in the sense that it is associated with Jacobi vector fields. This point requires some unpacking: we will first note that the derivative of the exponential map---with the exponential map being the inverse of the covariant derivative of the world function---is associated with a Jacobi vector field before then showing that the derivative of the world function itself is likewise associated with a Jacobi vector field. To start, consider a 1-parameter family of geodesics as given by $s: [a, b] \times (-\epsilon, \epsilon) \rightarrow T_{x'} M, (t, u) \mapsto tX + utY$ (where $t$ is the `time' parameter, $u$ is the `deviation' parameter and $(-\epsilon, \epsilon)$ denotes the range of the deviation parameter) composed with the exponential map $\exp_{x'}: U \subset T_{x'} M \rightarrow M$ so that
    $H = \exp_{x'} \circ s: [a, b] \times (-\epsilon, \epsilon) \rightarrow M, (t, u) \mapsto \exp(tX + utY)$ and $H(t, u=\text{const})$ is a geodesic for any sufficiently small $|u|$. 
    We can then note that the variation vector $V(t) := \frac{\partial}{\partial u} H\Bigr|_{u=0}$ relative to the geodesic $H(t, 0)$ fulfils the Jacobi equation 
\begin{equation}
\frac{d^2}{dt^2} V(t) = R \left( \frac{d H(t, 0)}{dt}, V(t) \right) \frac{d H(t, 0)}{dt}
\end{equation}
with Ricci tensor $R$ and the initial conditions $V(0) = 0$ and $\frac{dV}{dt} (0) = Y$.
Clearly then, the differential of the exponential map $d (\exp_{x'}) (tX):  T_{tX} \left( T_{x'} M \right) \cong T_{x'} M \rightarrow T_{\exp_{x'}(tX)} M,  T_{x'} M \ni {t Y} \mapsto V(t)$ is associated directly to a Jacobi field and thus geodesic deviation structure.\footnote{See, for instance, \citepbjps{Wilkins} for more details.}
As the differential of the inverse of a bijective function $f$ is just the inverse of the differential $df$, so is then the differential of its inverse, $d \sigma^{a'}_{x'}(x) = d \sigma^{a'}_{x'}(x(t,u))$.

Finally, in preparation for the following, let us furthermore define $\sigma^{a}(x, x'):= \nabla^{a} \sigma(x, x')$; as well as $\sigma\indices{_{a b'}} := \nabla_a \nabla_{b'} \sigma(x, x')$, $\sigma\indices{_{a' b}} := \nabla_{a'} \nabla_{b} \sigma(x, x')$, $\sigma\indices{_{a b}}  := \nabla_{a} \nabla_{b} \sigma(x, x')$, and $\sigma\indices{_{{a'} {b'}}} := \nabla_{a'} \nabla_{b'}
\sigma(x, x')$ where all expressions are invariant under swapping of indices  \citepbjps[\S3.2]{poisson2011motion}.

\subsubsection{Riemann normal coordinates}

\label{sec:riemannnormalcoordinates}

Relative to the fixed base point $x'$ and the tetrad $e\indices{_{I}^{a'}}$, the world function can be used to assign coordinates to a neighbouring point $x$ of form
\begin{equation}
\hat{x}^I = e\indices{^{I}_{a'}} \sigma^{a'}(x, x').
\end{equation}
Notably, the RNCs are a down-projection of the (coordinate-independent) bitensorial object $\sigma^{a'}(x, x')$ by means of the duals to the tetrad vectors $e\indices{_{I}^{a'}}$ at the base point $x'$. In particular, one can easily switch back and forth between the coordinate-dependent and coordinate-independent viewpoint relative to the base point $x'$. 
This demonstrates clearly that RNCs have directly underlying coordinate-independent geometric structure: this coordinate-independent geometric structure is specified in terms of the derivative of the world function at base point $x'$ (see \S \ref{sec:worldfunction}). 

One often sees RNCs constructed at the base point $x'$ via (i) the exponential map $\exp_{x'}: T_{x'} M \rightarrow \mathcal{N}(x') \subset M$, $v \mapsto \gamma_{\nu} (1)$ where $\gamma_v$ is the unique geodesic for which $\gamma_v (0) = x'$ and for which $\gamma'_v (0) = v$---with uniqueness guaranteed from requiring $\mathcal{N}(x')$ to be a normal neighbourhood---and (ii) the canonical isometric isomorphism $E$ between $(T_{x'} M, g_{x'})$ and $(\mathbb{R}^n, \eta)$, i.e., \[E: T_{x'} M \rightarrow \mathbb{R}^n, v \mapsto \left(g_{x'} (v, e_1), \dots, g_{x'} (v, e_n) \right)^T\] with $g_{x'}(e_I, e_J) = \eta_{IJ}$ and $T$ denoting the transpose; namely, then $\phi = E \circ \exp_{x'}^{-1} : M \supset \mathcal{N}(x') \rightarrow \mathbb{R}^n$ assigns to a point $x \in \mathcal{N}(x')$ the normal coordinates $(\hat{x}^1, \dots, \hat{x}^n)$.  
%relative to the orthonormal frame $e_I$ for $\mathbb{R}^n$. 
But this construction is in fact just the one above: the inverse to exp is generated by the world function as $\sigma^{a'}_{x'} = \exp_{x'}^{-1}: \mathcal{N}(x') \subset M \rightarrow T_{x'}M, x \mapsto \sigma^{a'}(x)$ (as explicated in \S \ref{sec:worldfunction}); $E$ is given by $(e^1, ...., e^n)$ with $e^{I}: T_{x'} M \rightarrow \mathbb{R}, v^{a'} \mapsto e^{I}_{a'} v^{a'}$. Thus, $\phi(x) = (e^{1}_{a'} \sigma^{a'}, \dots, e^{n}_{a'} \sigma^{a'})=(\hat{x}^1, \dots, \hat{x}^n)$.

Within GR, of paramount importance is that RNCs provide an asymptotic expansion of the metric:
\begin{equation}
g_{IJ} = \eta_{IJ} - \frac{1}{3} R_{ILJK} \hat{x}^L \hat{x}^K + O(x^3)
\end{equation}
where $R_{ILJK} : = R_{a'b'c'd'} e\indices{_{I}^{a'}} e\indices{_{L}^{b'}} e\indices{_{J}^{c'}} e\indices{_{K}^{d'}}$. Considering that $[R_{ILJK}] = \text{length}^{-2}$ and $[\hat{x}^I] = \text{length}$, one obtains upon explicitly accounting for the relevant measurement scales (by which we mean writing each dimensionful geometric object as a dimensionful scalar multiplied by a dimensionless geometric object---see \citepbjps{mana2021dimensional}) that 
\begin{equation}
\label{eq:asymptotic}
g_{IJ} = \eta_{IJ} - \delta^2 \frac{1}{3} \tilde{R}_{ILJK} \tilde{x}^L \tilde{x}^K + O(x^3)
\end{equation}
with $\delta : = \frac{l_{\text{probing}}}{l_{\text{curvature}}}$, i.e. the ratio of characteristic lengths as set by the experimental apparatus used for determining short and long scales respectively (think of two rods, one used for measuring the probing length and one for measuring the curvature length---then, they have length $l_\mathrm{probing}$ and $l_\mathrm{curvature}$, respectively); $\tilde{R}_{ILJK}$ and $\tilde{x}^L$ are both dimensionless.\footnote{Such an explicit rewriting of an expansion series in terms of the relevant measurement scales is common in physics. See for instance \citepbjps[§2.2]{Manohar} for a pedagogical example in the context of the multipole expansion---the latter of which we discuss further below.}\textsuperscript{,}\footnote{\label{footnote:kothawala} This is not to say that this is the only well-behaved asymptotic expansion possible: a liberated sense of Riemann normal coordinates around maximally symmetric spacetimes more generally (and not just Minkowski spacetime) is given by \citetbjps{hari2020normal} in their equation (25). As the authors themselves point out, this expansion is not as purely adapted to the inertial structure as the Riemann normal coordinates (the rest frame is `non-inertial').}

In fact, given that RNCs are really just down-projections of geometric structure encoded in the $\sigma^{a'}$, we can also provide the expansion of the metric in terms of the derivative of the world function alone (i.e., prior to a specific down-projection via a tetrad at the base point of the expansion):
\begin{equation}
g_{a'c'} = \eta_{a'c'} - \delta^2 \frac{1}{3} R_{a' b' c' d'} \sigma^{b'} \sigma^{d'} + O( (\nabla \sigma)^3).
\end{equation}
This shows in particular that the expansion parameter $\delta$ is independent of the specific tetradic down-projection and thus intrinsic to the geometrical structure of the general relativistic model under consideration.

It is worth stressing that expanding `around a point' is not to be understood literally. Based upon what we said about the representational nature of a point in \S\ref{sec:labframe}, we should really think of the expansion point as standing for some object of completely negligible extent in the modelling context; the effective physics in the modelling context is to be described in the normal coordinates around it with the relevant metric for instance described by \eqref{eq:asymptotic}. The individual points in the neighbourhood are not to be seen as basic, dimensionless entities but ultimately as (albeit not the only\footnote{Clearly, the expansion \eqref{eq:asymptotic} itself provides clear restrictions on the validity of the effective physical description.}) reminders of the resolutional restriction in the modelling context (cf.~how the moon can be modelled as a point particle when describing its orbit around the sun at a certain effective scale, with the effective physics being described through an elliptic curve).

\subsubsection{Other normal coordinates}

\label{sec:ferminormalcoordinates}

One can extend one's notion of RNCs by extending the restrictions from a point $x'$ to a larger sub-neighbourhood. Fermi normal coordinates, for instance, extend the strict conditions of the RNCs for a point to a timelike curve. This gives the Fermi normal coordinates a natural interpretation as the frame of an observer, and that of an inertial observer specifically if the curve is also a geodesic. 

For all these different adaptation and construction schemes beyond RNCs, we note that they---just as RNCs themselves---are related to the geodesic structure through second-order contact. The remarkable aspect of RNCs is that they are, however, the immediate `down-projection' of the world function and thus of the geodesic deviation structure (see again \S\ref{sec:riemannnormalcoordinates}); they are therefore the coordinates that are immediately connected to the geodesic deviation structure. 

\subsection{The scale-relative perspective}

\label{sec:EFTlikeaspect}

Above, we sketched how the inter-related formal tools of geodesic deviation and normal coordinates are used to spell out the effective emergence of SR within GR in a particular regime (i.e.,~they provide an asymptotic expansion of the metric). In this subsection, we explain how this way of modeling the `effective physics of SR' is just taking a scale-relative perspective, widely standard in modern physics (for instance, in the context of effective field theories (EFTs)), \emph{within the differential geometric formalism of GR}. Having done so, we will be in a position to understand why such an expansion around Minkowski spacetime is singled out over that around other backgrounds.

This scale-relative perspective we advocate has much which one could also characterise as `EFT-like' (such as the involvement of reasoning in terms of attractor `fixed points'). Note that by `EFT-like', we do not mean the literal application of the EFT formalism (familiar from high energy and condensed matter physics) to the theory of GR (though that can be done---see \citepbjps{donoghue}, \citepbjps{goldberger2006effective}, \citepbjps{porto2016effective}). Rather, by an EFT-like account of the emergence of SR, we mean an account whose conditions, motivations, rough structural features, and explanatory strategy bear a strong analogy to what we typically think of as EFT models of physics.\footnote{One clear difference between the traditional discussion of EFTs, RG flow, and attractor fixed points on the one hand, and our discussion on the other, is that the former typically extracts the flow from the formal dynamics (e.g. the Hamiltonian or Lagrangian) of the system under consideration, whereas the system that we consider here is formally `kinematical' and is thus not rich enough to support the aforementioned extraction procedure. Those differences notwithstanding, we still maintain that the analogy here is a helpful one---as, for example, we can still have a notion of `flow' based upon the changing of length scales. Our thanks to a reviewer for pushing us on this.}

In order to motivate our scale-relative perspective on the formalism of GR, let us begin by recalling that physics always concerns the description of subsystems that are in some sense isolated from external influences; furthermore, we are not in the business of describing perfectly precise quantities (e.g. real-number valued positions) of these subsystems but only such quantities up to a precision that makes sense relative to the physics of the subsystem and the environment (including the physics of the measuring apparatus). The stability or isolation of these subsystems implies that there is a separation of scales---the scale(s) of the subsystem from the scale(s) of other physical phenomena from which the subsystem is decoupled.  

All this is just part and parcel of the everyday practice of physics. Scale-relative models of physics try to go further by representing this separation of scales explicitly as part of the model, and using it to define a small expansion parameter $\delta$ (often called the `power counting parameter') such that quantities can be (formally) calculated to some order $n$ in the expansion parameter $\delta$, where each higher-order term represents a correction to the previous term and thus a more precise description. This is of course not to say that physical (as opposed to formally-defined) quantities have a precision that can be improved arbitrarily, but rather that explicitly parametrizing the precision by the order of $\delta$ gives us a powerful form of idealization and thereby a means of describing quantities up to some physically meaningful finite precision.  

In order to illustrate these features, consider a very simple textbook example: the multipole expansion in electrostatics. In this scenario, there are two scales in play: (i) the typical spacing $a$ between the charges, which plays the role of the short-distance scale; and (ii) the distance $R$ between the localized charge distribution and some distant observer $O$.
If the scales are widely separated, \emph{viz}.,~$R \gg a$, then $\delta := a / R$ plays the role of the small expansion parameter. 
The electrostatic potential $V(r)$ can then be expanded in terms of $\delta$ by means of the familiar multipole expansion
\begin{equation}\label{multipole}
V(r) = \frac{1}{r} \sum_{l, m} c_{lm} \delta^l Y_{lm}(\Omega),
\end{equation}
where $c_{lm}$ are dimensionless coefficients and the $Y_{lm}(\Omega)$ are spherical harmonics. 
The powerful form of idealization referred to above arises when we formally take the limit in which the scales are `infinitely' separated, i.e. $\delta \rightarrow 0$, in which case $V(r)$ is just given by the term for a point charge, or monopole. 
In more contemporary EFT-parlance, the monopole plays the role of an `attractive fixed point' for the space of models parametrized by (\ref{multipole}), since all these models flow to the monopole as $\delta \rightarrow 0$.

Similarly, in the practice of GR, we also come across such scale-relative thinking in terms of an expansion parameter $\delta = \frac{l_{\text{probing}}}{l_{\text{curvature}}}$ all the time---for example:
\begin{description}
    \item[Approximate Killing vectors in curved spacetime:] \citetbjps{Jacobson1995} achieves his famous reformulation of the field equations as thermodynamic equations of state by assuming an approximate Killing observer as well as an approximate Rindler observer. The feasibility of the assumption of such observers is controlled by $\delta$. That this control is not merely superficial cosmetics of rigour becomes clear in a follow-up work by \citetbjps{JacobsonEssay}, in which he ponders on the validity of such an assumption, allowing him to argue that only GR but not higher-order corrections can be taken to have a meaningful thermodynamic reinterpretation in the spirit of the 1995 article.\footnote{What is common to many instances of working with approximate Killing symmetries in the literature is that they rarely specify an exact notion, therefore leaving it open to the (unacquainted) reader to assess whether there is really a precise and preferred(!) sense in which the notion can be fleshed out. Such a sense is provided by the account of \citetbjps{Harte}, which we will present in \S\ref{sec:Harte}.}
    \item[Geometrical optics-limit in curved spacetime] The high-frequency limit for electromagnetic waves (or waves more generally) in curved spacetime involves the idealisation of letting $\delta$ go to zero \citepbjps[\S22.5]{Misner:1973prb}.\footnote{See for instance \citepbjps{LinnemannRead} for discussion of the geometrical optical limit in this spirit.}
\end{description}

We should like to emphasize that yet another instance of such scale-relative reasoning is the asymptotic expansion of the metric around Minkowski in GR with $\delta$ as defined in the previous section through \eqref{eq:asymptotic}. Here, it is the Minkowski metric term $\eta_{IJ}$ which plays the role of an attractive fixed point, in the sense that the models parametrized by \eqref{eq:asymptotic} flow towards this term as $\delta \rightarrow 0$ (furthermore, it is of course true that Minkowski spacetime would be self-similar under such a flow). We note that the value of $\delta$, and therefore the size of the neighbourhood both relative to the probing scale and relative to the background curvature scale, genuinely controls the approximation.\footnote{This is not a property possessed by the analysis of \citetbjps{fletcher2022local}, which works with a fiducial Riemannian metric. More on this in \S\ref{sec:fletcher}.}

So, the geodesic structure of GR bakes an EFT-like attractor fixed point into the theory. This also clarifies the sense in which Minkowski spacetime is special: of course, some other non-Minkowskian spacetime can approximate any spacetime in some regime (as \citetbjps{fletcher2022local} argue---see \S\ref{sec:fletcher}). However, for the asymptotic expansion in terms of `pure' geodesic deviation structure in the sense of \eqref{eq:asymptotic},\footnote{As opposed to the the expansions by \citetbjps{hari2020normal} mentioned in footnote \ref{footnote:kothawala}.} only the approximation to Minkowski spacetime becomes ever more accurate for ever smaller $\delta$.

\section{The local validity of special relativity from approximate Killing fields}
%- (Hawking and Ellis?)

\label{sec:localvalidity2}

RNCs and the associated coordinate-independent geodesic deviation structure capture completely what is meant by effective special relativistic geometry and/or behaviour. But just as one often characterises special relativistic geometry and/or behaviour in terms of symmetry structure (the Killing vectors of Minkowski spacetime), one might also want to account for the effective special relativistic geometry and/or behaviour in terms of approximate Killing vectors.

For instance,  \citetbjps{HawkingEllis} make use of an approximate Killing vector setup in RNCs to argue for the existence of locally approximately conserved currents in curved spacetime even in the absence of global Killing symmetries:

\begin{quote}
If the metric is not flat there will not, in general, be any Killing vectors and so the [...] integral conservation laws will not hold. However, in a suitable neighbourhood of a point $q$ one may introduce normal coordinates $\{x^{\alpha}\}$. Then at $q$ the components $g_{ab}$ of the metric are $e_a \delta_{ab}$ (no summation), and the components $\Gamma^a_{bc}$ of the connection are zero. One may take a neighbourhood $\mathcal{D}$ of $q$ in which the $g_{ab}$ and $\Gamma^a_{bc}$ differ from their values at $q$ by an arbitrarily small amount; then the [covariant derivatives of flat spacetime Killing vectors] $\underset{\alpha}{L}{}_{(a; b)}$ and $\underset{\alpha\beta}{M}{}\indices{_{(a; b)}}$ will not exactly vanish in $\mathcal{D}$, but will in this neighbourhood differ from zero by arbitrarily small amount. Thus \[\int_{\partial \mathcal{D}} \underset{\alpha}{P}{}^b d\sigma_b \quad \text{and} \quad \int_{\partial \mathcal{D}} \underset{\alpha \beta}{P}{}^b d\sigma_b\] [where $P^a = T^{ab} K_b$, and $K_b$ stands for the flat spacetime Killing vector] will still be zero in the first approximation; that is to say, one still has approximate conservation energy, momentum and angular momentum in a small region of space-time. \citepbjps[pp.~62-63]{HawkingEllis}
\end{quote}

In order to demonstrate that the approximate symmetry account from the RNC perspective has a clear coordinate-independent counterpart, slightly more work needs to be put in than was necessary for showing that the RNC perspective corresponds to coordinate-independent geometric structure. Fortunately, this work has  been undertaken by \citetbjps{Harte}, whose results we present and contextualise over the course of this section.\footnote{For a recent application of \citepbjps{Harte} to philosophical issues regarding energy conservation and causation in GR, see \citepbjps{MurgueitioRamirezForthcoming-MURCAT-11}.} In \S\ref{sec:Harte}, we introduce Harte's constructions; in \S\ref{sec:approximate}, we demonstrate how to think about local approximate symmetries using this machinery; in \S\ref{sec:EFTlike2}, we once again highlight the scale-relative nature of these considerations.

\subsection{Harte's construction}

\label{sec:Harte}

In this subsection, we will see via the work of \citetbjps{Harte} how an approximate Killing field $\psi_K^{d}$ is immediately constructed from geodesic structure and the constraint that at a base point $x'$ one has Killing symmetry:\footnote{We will see in the course of the section why such a constraint can be imposed at any point.} 
\begin{equation}
\label{eq:approximateKilling}
\mathcal{L}_{\psi_K^{d'}} g_{b' c'} = 0,
\end{equation}
i.e. $\psi_K^{d'}$ (that is, $\psi_K^{d}$ at $x'$) is exactly Killing.

In fact, following \citetbjps{Harte}, we first introduce the more general notion of an approximate `affine colineation field' $\psi^d$ that is immediately associated with geodesic structure and an affine colineation constraint at the base point $x'$, i.e. the constraint that at a base point $x'$ one has affine colineation: 
\begin{equation}
\label{eq:approximateColineation}
\nabla_{a'} \mathcal{L}_{\psi^{d'}} g_{b' c'} = 0.
\end{equation}
i.e. $\psi^{d'}$ (that is, $\psi^{d}$ at $x'$) is an exact affine colineation.\footnote{The existence of an affine colineation field at a point can be read as the statement that the Levi-Civita connection is conserved at that point. Clearly, any Killing field at point is also an affine colineation at that point.}

Notably, an approximate Killing field is automatically an approximate affine colineation. As we will learn in a second step, these approximate affine colineation fields (and thus in particular approximate Killing fields) are geodesic deviation fields themselves (also known as Jacobi fields).

The basic strategy of Harte is to use the exponential map and its inverse to induce uniquely the values for $\psi^d$ at $x \neq x'$ from deformations of the Killing vector $\psi^{d'}$ on the tangent space of $x'$.
Importantly, there are two contributions to a deviation field of this kind: first, linear deformations of vectors $X$ in the tangent space of $x'$ correspond, via the exponential map relative to $x'$, to deformations of $x'$ to a point $x$ by a path segment $\delta x$ on the manifold.\footnote{We provide the technical definitions and further details for the two forms of deviations considered here in Appendix \ref{appendix:harte}.} The path segment $\delta x$ in turn corresponds, via the inverse of the exponential map relative to point $x$, to a (deviation) vector at that point $x$. This is the first contribution to $\psi^d$. And second, changes of the base point $x'$ itself that do not affect $x$ can be taken to induce a transformation at the point $x$, and thus also a second contribution to $\psi^d$, if the change of $x'$ is re-interpreted as an active transformation which leaves $x'$---after all the center of reference---the same but changes $x$.  Bringing both forms of deviations together, the full set of transformations is \begin{equation}\psi^a = H\indices{^{a}_{a'}} \left( \sigma\indices{^{a'}_{b'}} A^{b'} - \sigma_{b'} B\indices{^{b' a'}} \right).\end{equation}
with $\psi^{a'} (x', x') = A^{a'}$ and $\nabla_{b'} \psi_{a'} = B_{b' a'}$. (In Minkowski spacetime specifically, $\psi^{\alpha} = A^{\alpha} + (x - x')^{\beta} B\indices{_{\beta}^{\alpha}}$.)

As mentioned at the beginning of this section, \emph{approximate} affine colineation fields $\psi^a$ are equivalently characterised as Jacobi fields; these have initial conditions $\psi^{a'}(x', x') = A^a$ and $B^{a'b'} = \nabla^{a'} \psi^{b'}(\gamma, \gamma)$, i.e.~$\psi^d$ fulfils the Jacobi equation (also known as the geodesic deviation equation)
\begin{equation}
\label{affinecolineation}
\sigma^{b} \sigma^{c} (\nabla_b \nabla_c \psi_a - R\indices{_{abc}^d} \psi_d) = 0.
\end{equation}
Here, the corresponding Jacobi propagators $H\indices{^a_{a'}} \sigma\indices{^a_{a'}_{b'}}$ and $H\indices{^a_{a'}} \sigma_{b'}$ are exactly the deformations as derived in the construction via the exponential map construction. 
(Exact affine colineations fulfil $\nabla_a \mathcal{L}_\psi g_{ab} = 0$---or rather the equivalent statement $\nabla_b \nabla_c \psi_a - R\indices{_{abc}^d} \psi_d = 0$---and are thus a \textit{special} subset of Jacobi fields.)
%\footnote{See the appendix of \cite{Harte} for a derivation.}

In sum, then, we obtain reassurance that the notion of approximate affine colineation \emph{qua} Jacobi field is a generalisation of exact affine colineations in terms of geodesic deviation and geodesic deviation structure alone. Note that the special case of picking $A$ and $B$ with $B$ antisymmetric as initial values for the geodesic deviation equation gives a geodesic deviation structure that \textit{only} generalises Killing symmetries (such that $\mathcal{L}_{\psi} g_{a'b'} = 0$), called \emph{approximate} Killing fields.

\subsection{Normal coordinates and approximate Killing symmetries}

\label{sec:approximate}

In this section, we investigate the extent to which normal coordinates stand in a special relation to approximate Killing vector fields. \emph{Prima facie}, one would expect the flow of tetrads along the integral curves of the approximate Killing vector fields to be in continuity with the flow of tetrads along the integral curves of Killing vector fields.

Consider first the behaviour of tetrads under exact Killing symmetries $\xi$ corresponding to rotation/boost. \citetbjps{chinea1988symmetries} derives that $\mathcal{L}_{\xi} e^I = \chi\indices{_{\xi}^I_J} e^J$ with $\chi_\xi$ an element of the Lie algebra $\mathfrak{so}(3,1)$. What is the situation for approximate Killing symmetries? If Killing symmetries rotate/boost tetrads, then approximate Killing symmetries should rotate/boost tetrads at least to zeroth order. In fact, and in analogy to Harte's approximation of the Lie derivative of the metric away from $x'$ \citepbjps[eq.~32]{Harte}, one obtains for the Lie derivative on the tetrad
\begin{equation}
\mathcal{L}_{\psi} e^I_a = \sigma\indices{^{a'}_{a}} \mathcal{L}_{\psi} e^I_{a'} + \sigma\indices{^{a'}_a} X^{c'} [\nabla_{c} \mathcal{L}_{\psi} e^I_{a}]_{a, c\rightarrow a', c'} + \frac{1}{2} \sigma\indices{^{a'}_{a}} X^{d'} X^{c'} [\nabla_{d} \nabla_{c} \mathcal{L}_{\psi} e^I_{a}]_{a, c, d\rightarrow a', c', d'} + \ldots,
\end{equation}
where an index limit of form $a \rightarrow a'$ denotes a point limit $x \rightarrow x'$ relative to that tensor index. %COMMENT_9
The zeroth order term is simply evaluated to $\chi\indices{_{\psi}^{I}_{J}} e\indices{^J_{a'}}$ given that $\psi$ is exactly Killing at $x'$. For further simplification, one uses that $\psi$ is an affine colineation at $x'$, i.e., the connection and Lie derivative commute.  The first order term then becomes $[\nabla_{c} \mathcal{L}_{\psi} e^I_{a}]_{a, c\rightarrow a', c'} = \mathcal{L}_{\psi} \nabla_{c'}  e^I_{a'}$.
Upon rewriting as in \S \ref{sec:normalcoordinates}, we obtain:
\begin{equation}
\mathcal{L}_{\psi} e^I_a = \sigma\indices{^{a'}_{a}} \mathcal{L}_{\psi} e^I_{a'} + \delta \cdot \sigma\indices{^{a'}_a} X^{c'} \mathcal{L}_{\psi} \nabla_{c'}  e^I_{a'} + \frac{\delta^2}{2} \sigma\indices{^{a'}_{a}} X^{d'} X^{c'} [\nabla_{d} \nabla_{c} \mathcal{L}_{\psi} e^I_{a}]_{a, c, d\rightarrow a', c', d'} + \ldots ,
\end{equation}
where $\delta = \frac{l_{\text{probing}}}{l_{\text{curvature}}}$ as before.\footnote{The Lie derivative acts on the tetrad conceived as a set of four vectors on spacetime---not conceived of as a set of four vectors in the internal lab frame. With respect to the latter, i.e., the lab frame index $I$ alone, it is clear that Lorentz transformations relate one tetrad frame to another. (That transformations between lab frames are exactly the Lorentz transformations can be seen by considering that (1) the transformation from tetrad $e\indices{_a^{\mu}}$ to $\tilde{e}\indices{_a^{\mu}}$ is given by $\tilde{e}\indices{_a^{\mu}} \tilde{e}\indices{^b_{\mu}}$---a matrix which fulfils `$\Lambda \Lambda \eta = \eta$' and thus counts as Lorentz transformation---; and (2) that a Lorentz transformation acting on a tetrad index $I$ results in another tetrad, i.e.~is a transformation between tetrads.)}

\subsection{The scale-relative perspective again}

\label{sec:EFTlike2}

We can illustrate the scale-relative perspective worked out before also upon the reconceptualisation of special relativistic physics in terms of its characteristic symmetry group---the Poincaré group---and its corresponding Killing symmetries: just as the asymptotic expansion in RNCs renders only the Minkowski metric self-similar (see \S\ref{sec:EFTlikeaspect}), the asymptotic expansion of the Lie derivative on tetrads renders only the exact Killing symmetries as having a self-similar effect.

Now, do all (approximate) Killing vectors one usually encounters for spacetimes effectively `align' with Harte's approximate symmetries upon some limit? In four dimensions they clearly do: four dimensional spacetime only allows for ten different approximate Killing vectors. These possibilities are exhausted by the full (approximate) symmetry groups dS$_4$, AdS$_4$, and Poincaré. Upon taking a group contraction tantamount to zooming out---letting the background curvature scale $R \rightarrow \infty$ while keeping the probing scale fixed, so $\delta \rightarrow 0$)---the only possibility remaining is the full symmetry group of Minkowski spacetime.\footnote{See \citepbjps{ayala2002group} and \citepbjps[§10]{enayati2022sitter} for accessible accounts.} Notably, the point generalises for all spacetime dimensions $n \geq 2$. Ultimately, this means that with respect to group contraction, any set of approximate Killing vectors is an approximation of the Poincaré Killing vectors---explicitly, the reasoning proceeds as follows:
\begin{enumerate}
    \item Any full set of approximate Killing vectors with control parameter $\delta$ is an approximation of a full set of exact Killing vectors (with exactness at the base point).
    \item In $d=4$, there are just three sets of exact Killing vectors: AdS, dS and Poincaré Killing vectors.
    \item dS and AdS symmetry groups are approximations with control parameter $\delta$ to the Poincaré symmetry group (group contraction with respect to $\delta \rightarrow 0$ results in the Poincaré symmetry group).
    \item Therefore, any full set of approximate Killing vectors with control parameter $\delta$ is ultimately an approximation of the Poincaré Killing vectors (with exactness at the base point).
\end{enumerate}

\section{Responses to some other views}

\label{sec:response}

In this section, we consider how the scale-relative perspective on the local validity of SR which we have offered above interacts with recent work on this issue by \citetbjps{fletcher2022local}\footnote{For earlier related work by these authors, see  \citepbjps{FletcherApprox} and \citepbjps{WeatherallDogmas}.} (\S\ref{sec:fletcher}) and \citetbjps{gomes} (\S\ref{sec:gomes}).

\subsection{On Fletcher and Weatherall}

\label{sec:fletcher}

\citetbjps{fletcher2022local} focus in their first article on geometric aspects of the recovery of SR, and discuss two interpretations of what the `local validity of SR' could mean which are of particular importance:\footnote{What \citetbjps{fletcher2022local} dub the `Literal interpretation' (every spacetime is locally isometric to Minkowski spacetime) and `Coordinate-chart interpretation, first pass' (according to which there are coordinates in which curvature vanishes) are straightforwarldy false explications of the local validity of SR (as \citetbjps{fletcher2022local} go on to point out themselves).}
\begin{quote}
\textbf{Tangent space interpretation:} ``The tangent space at a point of spacetime is, or is somehow equivalent to, Minkowski spacetime.'' \citepbjps[p.~7]{fletcher2022local}
\end{quote}
and\footnote{`Coordinate Chart Interpretation, second pass', in the words of \citetbjps{fletcher2022local}.}
\begin{quote}
\textbf{Normal coordinate interpretation:} ``At any point of any relativistic spacetime (or along certain curves), local coordinates may be chosen so that, at that point (or along that curve), (a) the components of the metric agree with the Minkowski metric in standard coordinates and (b) all Christoffel symbols vanish." \citepbjps[p.~10]{fletcher2022local}
\end{quote}

Our line in this article has been that both the tangent space interpretation and the normal coordinate interpretation are integral parts of cashing out what one means by the local validity of SR in GR:\footnote{We do not think it is \emph{de facto} justified to keep separate the notions of local special relativistic \textit{geometry} and local special relativistic dynamical \textit{behaviour}: the effective understanding of local special relativistic geometry (as explicated in terms of geodesic deviation structure and approximate Killing symmetries) is understood directly with respect to how matter moves, both kinematically as well as at the level of the equations of motion.} (i) the tangent space at a point of spacetime is equivalent to Minkowski spacetime in the sense of acting as the `fixed point' around which a preferred small parameter expansion takes place, and (ii) the normal coordinate interpretation fleshes out concretely that preferred small parameter expansion. By appreciating furthermore that (iii) the expansion is to be understood from an effective field theory point of view, one indeed obtains all-together a precise and preferred sense in which GR is locally special relativistic.

In any case, Fletcher and Weatherall's most charitable take on the standard story is as either claiming that (A) the existence of normal coordinates expresses certain facts about local geometry, or that (B) there is an interpretive claim about idealized measurement apparatuses of natural motions that somehow gets bundled up with normal coordinates. But they maintain that these claims are unsatisfactory for making sense of the local validity of SR.

Let's consider these in turn. With regard to (A), Fletcher and Weatherall have the understandable worry that it is  unclear how the `form' of such coordinates can tell us something special about the metric and covariant derivative given that there are also always coordinates in which the Christoffel symbols do not vanish. The missing piece of the puzzle here is that RNCs correspond directly to the coordinate-invariant geometric structure of geodesic deviation (see \S \ref{sec:worldfunction})---other coordinate systems in general lack such an intimate connection to geometric structure.\footnote{In \emph{specific} spacetimes (such as spacetimes with rich isometry groups), there can of course be other coordinate systems which are very well-adapted to geometric structure.} With regard to (B), their point is that any such claim goes ``far beyond any facts about curvature, local or otherwise" \citepbjps[p.~1]{fletcher2022local}. But this going `far beyond' is simply to interpret the formalism as it is used in theorising---and that is provided by a scale-relative perspective, or so we have argued above. From a scale-relative point of view, the local validity of SR cannot simply be reduced to local facts about geometry: SR is a physical theory which must (partly) be understandable in terms of the local behaviour of apparatuses of some kind. Concretely, the connection between apparatus behaviour and normal coordinates is provided by the fact that the validity of normal coordinates is controlled by the scale size relevant for probing with apparatuses etc.~\emph{vis-à-vis} the background curvature scale.

One of the central goals of \citetbjps{fletcher2022local} is to make manifest whatever genuine content exists in the notion of the `local vaidity of SR' by providing a clear notion/interpretation of `locally flat'. For this, they introduce a metric distance function on the set of tensors under recourse to a fiducial Riemannian metric on the spacetime manifold, making the set of tensors thereby pointwise a metric space.\footnote{Concretely, for a neighbourhood $U \subset M$ with compact closure, choose any Riemannian metric $h_{ab}$. Relative to $h_{ab}$, define a norm on covariant tensors $f_{a_1 ... a_n}$ at a point by $|f|_h = |h^{a_1 b_1} ... h^{a_n b_n} f_{a_1 ... a_n} f_{b_1 ... b_n}|^{1/2}$; then define distance between $f:=f_{a_1 ... a_n}$ and $f':=f'_{a_1 ... a_n}$ as 
\begin{equation}
d_U (f, f'; h, k) = \text{max}_{j \in \{0, ..., k\}} \text{sup}_U |\nabla^j |f-f'|_k
\end{equation}
where $\nabla$ is the Levi-Civita derivatice with respect to $h_{ab}$.} This leads to (i) a metric space continuity notion of approximation of form. The latter in turn induces (ii) a notion of approximate symmetry.
Together, (i) and (ii) are used to articulate a `non-specialness' argument concerning their corresponding notion of `effective/approximate' Minkowski spacetime:
on this notion of approximate symmetry, a Lorentzian metric is not merely locally approximately the Minkowski metric (as follows from their Theorem 1) but really approximately any other metric (see their Theorem 5).

Of course these results are technically correct, and indeed in our view they are perfectly compatible with the scale-relative perspective offered in this article: it is one thing to understand the status of Minkowski spacetime as a fixed point attractor (as done above); it is another to develop (\emph{\`{a} la} \citetbjps{fletcher2022local}) a formal framework within which to cash out the notion of the closeness (or otherwise) of two metrics. That said, we worry that to focus \emph{exclusively} on geometric results such as Fletcher and Weatherall's Theorem 5 might lead one to downplay or underestimate the physical significance of Minkowski spacetime within the framework of GR. One way in which this is borne out is that the control parameter for an approximative expansion in terms of a fiducial metric is \emph{prima facie} without physical significance; this stands in contrast to the asymptotic expansion as given on the scale-relative account provided above, in which  $\delta$ keeps track of the deviation from the intrinsic geometrical structure of the general relativistic spacetime model at play.

\subsection{On Gomes}

\label{sec:gomes}

In direct response to \citetbjps{fletcher2022local}, \citetbjps{gomes} separates three senses of the local validity of SR in GR:
\begin{enumerate}
    \item At any $p \in M$, the tangent space is Minkowskian.
    \item At any $p \in M$, RNCs exist.
    \item In the relevant circumstances, geodesic deviation is negligable.
\end{enumerate}
Here, (1) is clearly the tangent space interpretation discussed above; (2) is clearly a version of the normal coordinate interpretation discussed above. 
Regarding (3), Gomes is surely correct to highlight this as an omission in the discussion offered by Fletcher and Weatherall. However, to stop here is to leave the story incomplete, for---as we have made clear in detail in the foregoing sections of this article---(1)--(3) are not independent notions of the local validity of SR in GR, but are in fact all related intimately to one another:
(1) underwrites (2), which itself is a way of expressing (3).

\section{Conclusions}

In this article, we have argued that there is a clear, natural, and ultimately uncontroversial sense in which GR can be said to be locally special relativistic; all that is required is to think of GR in terms of a scale-relative perspective, rather than as a mere set of Lorentzian geometries. 

Our position not only is consistent with what leading practitioners of GR have written on this topic (see e.g.~\citepbjps{WeinbergFirstEdition}, \citepbjps{HawkingEllis}, \citepbjps{PoissonWill}), but also takes seriously what philosophers of science have learned from numerous case studies of the empirical interpretation of a physical theory: said interpretation depends ultimately on bringing formalism into contact with relevant measurement scales, which are \emph{de facto} classical; for this to happen, objects and/or functional roles played by objects in that theory must be associated with (or at least accounted for in) objects and/or functional roles played by objects in a classical measurement theory, or with objects and/or functional roles played by objects in some intermediate theory.\footnote{Recently, similar points have been made by \citetbjps{Grimmer} in the context of measurements in QFT.}
Mere association, however, is insufficient: one must also account for the regime of validity of the approximation. For example, the role played by a coherent state in a quantum theory should be  linked effectively to the role played by a configuration in a classical theory---but only for the regime in which the state actually is coherent.

The same thinking underlies the account of the local validity of SR in GR which we have presented in this article: the local spacetime structure is to be linked effectively to a special relativistic---and ultimately a non-relativistic---structure; only if some expansion parameter is used can we thereby get an estimate of how good the measurement is according to GR.
From this point of view---and given that GR is empirically relevant---it is not surprising that there exists an asymptotic series of the metric with the Minkowski metric as `base point', or that a general relativistic spacetime has approximately and locally the symmetries of Minkowski spacetime.

\begin{appendices}

\section{Appendix}

\subsection{Harte's construction of the deviation field}

\label{appendix:harte}

Importantly, there are two contributions to a deviation field $\psi^a$ of the kind presented by \citetbjps{Harte}. First: linear deformations of vectors $X$ in the tangent space of $x'$ correspond, via the exponential map relative to $x'$, to deformations of $x'$ to a point $x$ by a path segment $\delta x$ on the manifold. The path segment $\delta x$ in turn corresponds, via the inverse of the exponential map relative to point $x$, to a (deviation) vector at that point $x$. This is the first contribution to $\psi^a$. We can understand this first contribution at a mere technical level: the goal is to state the deformation at $x$, denoted by the tangent vector $\psi^a$, in terms of a tangent vector at $x'$; to achieve this goal, define the linear deformation
\begin{equation}
\label{tangent}
X^{a'} \rightarrow X^{a'}(x, x') + \epsilon \Psi^{a'} 
\end{equation}
on the tangent space $T_{x'}M$ with $\Psi^{a'} := B\indices{_{b'}^{a'}} X^{b'}$. These transformations are  associated directly with a transformation $x \rightarrow x + \delta x$ on the manifold $M$ via the exponential map $\label{exp} x = \exp_{x'} X.$ The inverse of the exponential map is given by \begin{equation}\label{expinverse}X_{a'} = - \sigma_{a'} (x, x')\end{equation} in terms of the world function. From Taylor expanding the inverse, we learn that $X^{a'}(x+\delta x, x') = -\sigma_{a'} (x + \delta x, x') = - \sigma_{a'} (x, x') - (\epsilon \psi^b) \sigma_{b a'}(x, x')$. By comparison with \eqref{tangent}, it follows immediately that
\begin{equation}
\label{transformation1}
B\indices{_{b'}^{a'}} X^{b'}(x, x') = \psi^b \sigma_{b a'}(x, x').
\end{equation}
Upon defining $H\indices{^{b}_{a'}} = \left[- \sigma\indices{^{a'}_b} (x, x') \right]^{-1}$, we finally get 
\begin{equation}
\psi^b = - H\indices{^{b}_{a'}} B^{b' a'} \sigma_{b'},
\end{equation}
where we replaced $X_b'$ by $-\sigma_b'$ (in virtue of \eqref{expinverse}). Note that $B_{b'a'} = \nabla_{b'} \psi_{a'}(x', x')$: $\nabla_{b'} \psi^{a'}(x', x') = \nabla_{b'} \left( -H\indices{^{a'}_{c'}} \sigma_{d'} B^{d' c'} \right) = B\indices{_{b'}^{a'}}$ where the last step follows from the facts that $\sigma_d'(x', x') = 0$ and $\nabla_{b'} \sigma_{d}' = \sigma_{b' d'}$. 

For the second contribution to the deviation field $\psi^a$, regard $X^{a'}(x, x' + \delta x')$ as the result of a `passive transformation' of the description of $x$ relative to $x'$ (characterised by the fact that the reference point $x'$ but not $x$ changes); from Taylor expanding, we learn that $X^{a'}(x, x' + \delta x') = -\sigma_{a'} (x, x') -  \sigma_{a'b'} A^{b'}$. One then looks for the corresponding `active transformation' by equating $X^{a'}(x, x' + \delta x')$ to $X^{a'}(x + \delta x, x')$ (characterised by the fact that $x$ but not $x'$ changes). We  know already that a transformation relative to a base point $x'$ on a neighbouring $x$ generally has the form $X^{a'}(x+\delta x, x') = -\sigma_{a'} (x + \delta x, x') = - \sigma_{a'} (x, x') - \psi^b \sigma_{b a'}(x, x')$. Using the posited equality between a `passive transformation'---the base point \textit{qua} reference system is moved but not the point to be described---and an `active transformation'---the point to which we refer via the base point is moved but not the base point---, $\psi^{a}$ of the latter conception can be determined to be $\psi^a = H\indices{^{a}_{a'}} \sigma\indices{^{a'}_{b'}} A^{b'}$. Note that clearly $A^{a'} = \psi^{a'}(x', x')$ (the `prefactor' to $A$ becomes an identity with the coincidence at $x'$; $B_{b'a'}(x', x') = 0$ for translations in any case).

Bringing these results together, the full set of transformations is \begin{equation}\psi^a = H\indices{^{a}_{a'}} \left( \sigma\indices{^{a'}_{b'}} A^{b'} - \sigma_{b'} B\indices{^{b' a'}} \right),\end{equation}
with $\psi^{a'} (x', x') = A^{a'}$ and $\nabla_{b'} \psi_{a'} = B_{b' a'}$.

\end{appendices}

\section*{Funding}

N.L.~acknowledges the support of the Swiss National Science Foundation as part of the project `Philosophy Beyond Standard Physics' (105212\_207951). J.R.~acknowledges the support of the Leverhulme Trust. N.T.~and J.R.~both acknowledge the support of NSF Grant 1971455 and JTF Grant 61521.

\section*{Acknowledgements}

For very helpful discussions we are grateful to Erik Curiel, Dominic Dold, Sam Fletcher, JB Manchak, Sebastián Murgueitio Ramírez, Liviu Nicolaescu, Eric Poisson, Jim Weatherall, and Clifford Will.  

\begin{flushright}
\emph{
  Niels Linnemann\\
  University of Geneva\\
  Department of Philosophy\\
  Genève, CH\\
  niels.linnemann@unige.ch
}
\end{flushright}

\begin{flushright}
\emph{
  James Read\\
  University of Oxford\\
  Faculty of Philosophy\\
  Oxford, UK\\
  james.read@philosophy.ox.ac.uk
}
\end{flushright}

\begin{flushright}
\emph{
  Nicholas J. Teh\\
  University of Notre Dame\\
  Department of Philosophy\\
  Notre Dame, IN, USA\\
  nteh@nd.edu
}
\end{flushright}

\bibliographystyle{bjps}
\bibliography{references}

\begin{thebibliography}{39}
\expandafter\ifx\csname natexlab\endcsname\relax\def\natexlab#1{#1}\fi
\expandafter\ifx\csname url\endcsname\relax
  \def\url#1{{\tt #1}}\fi
\expandafter\ifx\csname urlprefix\endcsname\relax\def\urlprefix{URL }\fi

\bibitem[{Ayala and Haase(2002)}]{ayala2002group}
Ayala, M.  and Haase, R. [2002]:  `Group contractions and its consequences upon
  representations of different spatial symmetry groups',
\newblock {\em arXiv preprint hep-th/0206037\/}.

\bibitem[{Ballmann(2015)}]{Ballmann}
Ballmann, W. [2015]:  `{C}RITICAL POINT THEORY OF THE ENERGY FUNCTIONAL ON PATH
  SPACES', .

\bibitem[{Brown and Read(2021)}]{BrownRead}
Brown, H.~R.  and Read, J. [2021]:  `The Dynamical Approach to Spacetime
  Theories',
\newblock in E.~Knox  and A.~Wilson (\emph{eds}), {\em The Routledge Companion
  to Philosophy of Physics\/}.

\bibitem[{Cardin and Marigonda(2004)}]{Cardin}
Cardin, F.  and Marigonda, A. [2004]:  `{Global World Functions}',
\newblock {\em Journal of Geometry and Symmetry in Physics\/}, {\textbf{
  2}\/}(none), pp. 1 -- 17.
\newline <https://doi.org/10.7546/jgsp-2-2004-1-17>

\bibitem[{Chamseddine and Mukhanov(2010)}]{mukhanov}
Chamseddine, A.~H.  and Mukhanov, V. [2010]:  `Gravity with de {S}itter and
  unitary tangent groups',
\newblock {\em Journal of High Energy Physics\/}, {\textbf{ 2010}\/}(3), pp.
  1--18.

\bibitem[{Chinea(1988)}]{chinea1988symmetries}
Chinea, F. [1988]:  `Symmetries in tetrad theories',
\newblock {\em Classical and Quantum Gravity\/}, {\textbf{ 5}\/}(1), pp. 135.

\bibitem[{Donoghue(1995)}]{donoghue}
Donoghue, J.~F. [1995]:  `{Introduction to the Effective Field Theory
  Description of Gravity}',
\newblock {\em arXiv preprint arXiv:9512024\/}.

\bibitem[{Einstein(1916)}]{Einstein1916}
Einstein, A. [1916]:  `{Die Grundlage der allgemeinen Relativitätstheorie}',
\newblock {\em Annalen Der Physik\/}, {\textbf{ 49}\/}(7), pp. 769--822
\newblock Translated as ``The Foundations of General Relativity'' in Lorentz et
  al., ``The Principle of Relativity'', Methuen and Company, Ltd. Reprinted by
  Dover Publications (1923).

\bibitem[{Einstein(1921)}]{EinsteinPainleve}
Einstein, A. [1921]:  `Letter to {P}ainlev\'{e}',
\newblock Vol.~12, Doc.~314 CPAE.

\bibitem[{Enayati \emph{et~al.}(2022)Enayati, Gazeau, Pejhan, and
  Wang}]{enayati2022sitter}
Enayati, M., Gazeau, J.-P., Pejhan, H.  and Wang, A. [2022]:  `The de {S}itter
  group and its representations: a window on the notion of de Sitterian
  elementary systems',
\newblock {\em arXiv preprint arXiv:2201.11457\/}.

\bibitem[{Fletcher(2020)}]{FletcherApprox}
Fletcher, S.~C. [2020]:  `Apprxomate local {P}oincar\'{e} spacetime symmetry in
  general relativity',
\newblock in C.~Beisbart, T.~Sauer  and C.~W\"{u}thrich (\emph{eds}), {\em
  Thinking About Space and Time\/}, Birkh\"{a}user.

\bibitem[{Fletcher and Weatherall(2023)}]{fletcher2022local}
Fletcher, S.~C.  and Weatherall, J.~O. [2023]:  `The Local Validity of Special
  Relativity, Part 1: Geometry',
\newblock {\em Philosophy of Physics\/}, {\textbf{ 1}\/}.

\bibitem[{Goldberger and Rothstein(2006)}]{goldberger2006effective}
Goldberger, W.~D.  and Rothstein, I.~Z. [2006]:  `Effective field theory of
  gravity for extended objects',
\newblock {\em Physical Review D\/}, {\textbf{ 73}\/}(10), pp. 104029.

\bibitem[{Gomes(2022)}]{gomes}
Gomes, H. [2022]:  ```{I}s spacetime locally flat?": a note', .
\newline <http://philsci-archive.pitt.edu/21311/>

\bibitem[{Grimmer(2023)}]{Grimmer}
Grimmer, D. [2023]:  `The {P}ragmatic {QFT} {M}easurement {P}roblem and the
  need for a {H}eisenberg-like cut in {QFT}',
\newblock {\em Synthese\/}.

\bibitem[{Hari and Kothawala(2020)}]{hari2020normal}
Hari, K.  and Kothawala, D. [2020]:  `Normal coordinates based on curved
  tangent space',
\newblock {\em Physical Review D\/}, {\textbf{ 101}\/}(12), pp. 124066.

\bibitem[{Harte(2008)}]{Harte}
Harte, A.~I. [2008]:  `Approximate spacetime symmetries and conservation laws',
\newblock {\em Classical and Quantum Gravity\/}, {\textbf{ 25}\/}(20), pp.
  205008.

\bibitem[{Hawking and Ellis(1973)}]{HawkingEllis}
Hawking, S.~W.  and Ellis, G.~F. [1973]: {\em The large scale structure of
  space-time\/},
\newblock Cambridge university press.

\bibitem[{Hetzroni and Read(2023)}]{HetzroniRead}
Hetzroni, G.  and Read, J. [2023]:  `How to Teach General Relativity',
\newblock {\em British Journal for Philosophy of Science\/}.

\bibitem[{Hildebrandsson(2021)}]{Uppsala}
Hildebrandsson, V. [2021]:  `The {H}amiltonian formulation of geodesics',
\newblock Digitala Vetenskapliga Arkivet
\newblock URL:https://uu.diva-portal.org/smash/get/diva2:1589204/FULLTEXT01.pdf
  (version: 2021-06).
\newline <https://uu.diva-portal.org/smash/get/diva2:1589204/FULLTEXT01.pdf>

\bibitem[{Iliev(2006)}]{Iliev}
Iliev, B. [2006]: {\em Handbook of Normal Frames and Coordinates\/},.

\bibitem[{Jacobson(1995)}]{Jacobson1995}
Jacobson, T. [1995]:  `Thermodynamics of spacetime: the Einstein equation of
  state',
\newblock {\em Phys. Rev. Lett.\/}, {\textbf{ 75}\/}(7), pp. 1260.

\bibitem[{Jacobson(2012)}]{JacobsonEssay}
Jacobson, T. [2012]:  `Gravitation and vacuum entanglement entropy',
\newblock {\em International Journal of Modern Physics D\/}, {\textbf{
  21}\/}(11), pp. 1242006.

\bibitem[{Knox(2013)}]{Knox2013-KNOESG}
Knox, E. [2013]:  `Effective Spacetime Geometry',
\newblock {\em Studies in History and Philosophy of Science Part B: Studies in
  History and Philosophy of Modern Physics\/}, {\textbf{ 44}\/}(3), pp.
  346--356.

\bibitem[{Lehmkuhl(2021)}]{lehmkuhl2021equivalence}
Lehmkuhl, D. [2021]:  `The equivalence principle (s)',
\newblock in {\em The Routledge companion to philosophy of physics\/},
  Routledge, pp. 125--144.

\bibitem[{Linnemann and Read(2021)}]{LinnemannRead}
Linnemann, N.  and Read, J. [2021]:  `Comment on ‘{D}o electromagnetic waves
  always propagate along null geodesics?’',
\newblock {\em Classical and Quantum Gravity\/}, {\textbf{ 38}\/}(23), pp.
  238001.

\bibitem[{Mana(2021)}]{mana2021dimensional}
Mana, P.~P. [2021]:  `Dimensional analysis in relativity and in differential
  geometry',
\newblock {\em European Journal of Physics\/}, {\textbf{ 42}\/}(4), pp. 045601.

\bibitem[{Manohar(2018)}]{Manohar}
Manohar, A.~V. [2018]:  `Introduction to Effective Field Theories', .

\bibitem[{Misner \emph{et~al.}(1973)Misner, Thorne, and
  Wheeler}]{Misner:1973prb}
Misner, C.~W., Thorne, K.~S.  and Wheeler, J.~A. [1973]: {\em {Gravitation}\/},
\newblock San Francisco: W. H. Freeman.

\bibitem[{Murgueitio~Ram\'{i}rez
  \emph{et~al.}(forthcoming)Murgueitio~Ram\'{i}rez, Read, and
  Paez}]{MurgueitioRamirezForthcoming-MURCAT-11}
Murgueitio~Ram\'{i}rez, S., Read, J.  and Paez, A. [forthcoming]:  `Causation
  and the Conservation of Energy in General Relativity',
\newblock {\em The British Journal for the Philosophy of Science\/}.

\bibitem[{Poisson \emph{et~al.}(2011)Poisson, Pound, and
  Vega}]{poisson2011motion}
Poisson, E., Pound, A.  and Vega, I. [2011]:  `The motion of point particles in
  curved spacetime',
\newblock {\em Living Reviews in Relativity\/}, {\textbf{ 14}\/}(1), pp.
  1--190.

\bibitem[{Poisson and Will(2014)}]{PoissonWill}
Poisson, E.  and Will, C.~M. [2014]: {\em Gravity: Newtonian, post-{N}ewtonian,
  relativistic\/},
\newblock Cambridge University Press.

\bibitem[{Porto(2016)}]{porto2016effective}
Porto, R.~A. [2016]:  `The effective field theorist’s approach to
  gravitational dynamics',
\newblock {\em Physics Reports\/}, {\textbf{ 633}\/}, pp. 1--104.

\bibitem[{Read \emph{et~al.}(2018)Read, Brown, and Lehmkuhl}]{read2018two}
Read, J., Brown, H.~R.  and Lehmkuhl, D. [2018]:  `Two miracles of general
  relativity',
\newblock {\em Studies in history and philosophy of science Part B: Studies in
  history and philosophy of modern physics\/}, {\textbf{ 64}\/}, pp. 14--25.

\bibitem[{Synge(1973)}]{Synge}
Synge, J. [1973]:  `The general theory of relativity',
\newblock {\em Hermathena\/}, pp. 57--71.

\bibitem[{Trautman(2006)}]{EinsteinCartan}
Trautman, A. [2006]:  `Einstein-{C}artan theory',
\newblock {\em arXiv preprint gr-qc/0606062\/}.

\bibitem[{Weatherall(2020)}]{WeatherallDogmas}
Weatherall, J.~O. [2020]:  `Two dogmas of dynamicism',
\newblock {\em Synthese\/}, {\textbf{ 199}\/}, pp. 253--275.

\bibitem[{Weinberg(1972)}]{WeinbergFirstEdition}
Weinberg, S. [1972]:  `Gravitation and cosmology: principles and applications
  of the general theory of relativity', .

\bibitem[{Wilkins(2005)}]{Wilkins}
Wilkins, D.~R. [2005]:  `A {C}ourse in {R}iemannian {G}eometry', .
\newline <https://www.maths.tcd.ie/~dwilkins/Courses/425/RiemGeom.pdf>

\end{thebibliography}

\end{document}